\documentclass[preprint,prl,aps,epsf,amssymb]{revtex4}
\usepackage{graphicx}
\begin{document}

\draft

\title{The alpha effect and its saturation in a turbulent swirling flow 
generated in the VKS experiment}

\author {\bf F. P\'etr\'eli$\text s^{(1)}$, M. Bourgoi$\text n^{(2)}$, L.
Mari$\text {\'e}^{(3)}$, J. Burguet$\text e^{(3,4)}$ A. Chiffaude$\text l^{(3)}$, F. Daviau$\text
d^{(3)}$, S. Fauv$\text e^{(1)}$, P. Odie$\text r^{(2)}$, J.-F. Pinto$\text
n^{(2)}$.}

\address{(1)  Laboratoire de Physique Statistique de l'Ecole Normale
Sup\'erieure, UMR CNRS 8550, 24 Rue Lhomond, 75231 Paris Cedex 05, France;
\\ (2) Laboratoire de Physique de l'Ecole Normale
Sup\'erieure de Lyon, UMR CNRS 5672, 47 all\'ee d'Italie, 69364 Lyon Cedex 07, France;
\\ (3) Service de Physique de l'Etat Condens\'e, Direction des Sciences
de la Mati\`ere, CEA-Saclay, 91191 Gif-sur-Yvette cedex, France
\\ (4) Present address: Departamento de F\'{\i}sica y Matem\'atica
Aplicada, Universidad de Navarra, E-31080 Pamplona, Spain}

\date{\today}


\begin{abstract}

We report the experimental observation of the $\alpha$-effect. It
consists in the generation of a current parallel to a magnetic
field $\vec{B}_0$ applied to a turbulent swirling flow of
liquid sodium. At low magnetic Reynolds number, $R_m$, we show
that the magnitude of the $\alpha$-effect increases like $R_m^2$ and
that its sign is determined by the flow helicity. It saturates and
then decreases at large $R_m$, primarily because of the expulsion
of the applied field $\vec{B}_0$ from the bulk of the flow.
We show how this expulsion is affected by the flow geometry
by varying the relative amplitudes of the azimuthal and axial flows.
\end{abstract}

\maketitle

It has been first proposed by Parker that "cyclonic eddies" in an
electrically conducting fluid may generate a current parallel
to an applied magnetic field $\vec{B}_0$ \cite{par54}. This effect,
called the ``$\alpha$-effect", has been understood on a more quantitative
basis by Steenbeck, Krause and R\"adler \cite{kra80} and Moffatt 
\cite{mof78} in the case of scale separation, {\it i.e.} when the
magnetic  field 
has a large scale component compared to the scale of the eddies. The 
$\alpha$-effect is a key mechanism of most astrophysical and 
geophysical dynamo models \cite{mof78,rob94} and is also involved in 
the two recent laboratory observations of self-generation of a 
magnetic field by a flow of liquid sodium: the ``Karlsruhe experiment"
\cite{kar00}, which is an $\alpha^2$-type dynamo and the ``Riga 
experiment" \cite{rig00} which may be understood as an 
$\alpha\omega$-type dynamo \cite{rob87}. These experiments, as well as 
the only direct experimental study of the $\alpha$-effect \cite{ste68},
involve flows with geometrical constraints that are chosen in order to 
maximize the efficiency of the dynamo effect (respectively the 
$\alpha$-effect). Several groups are now trying to achieve 
self-generation of a magnetic field in turbulent flows without, or 
with less geometrical constraints, in order to study situations 
that are closer to astrophysical or geophysical models 
\cite{cargese}. It is thus of primary interest to study the 
$\alpha$-effect in such fully developed turbulent flows. 

\begin{figure}
\centerline{ \includegraphics[width=10cm]{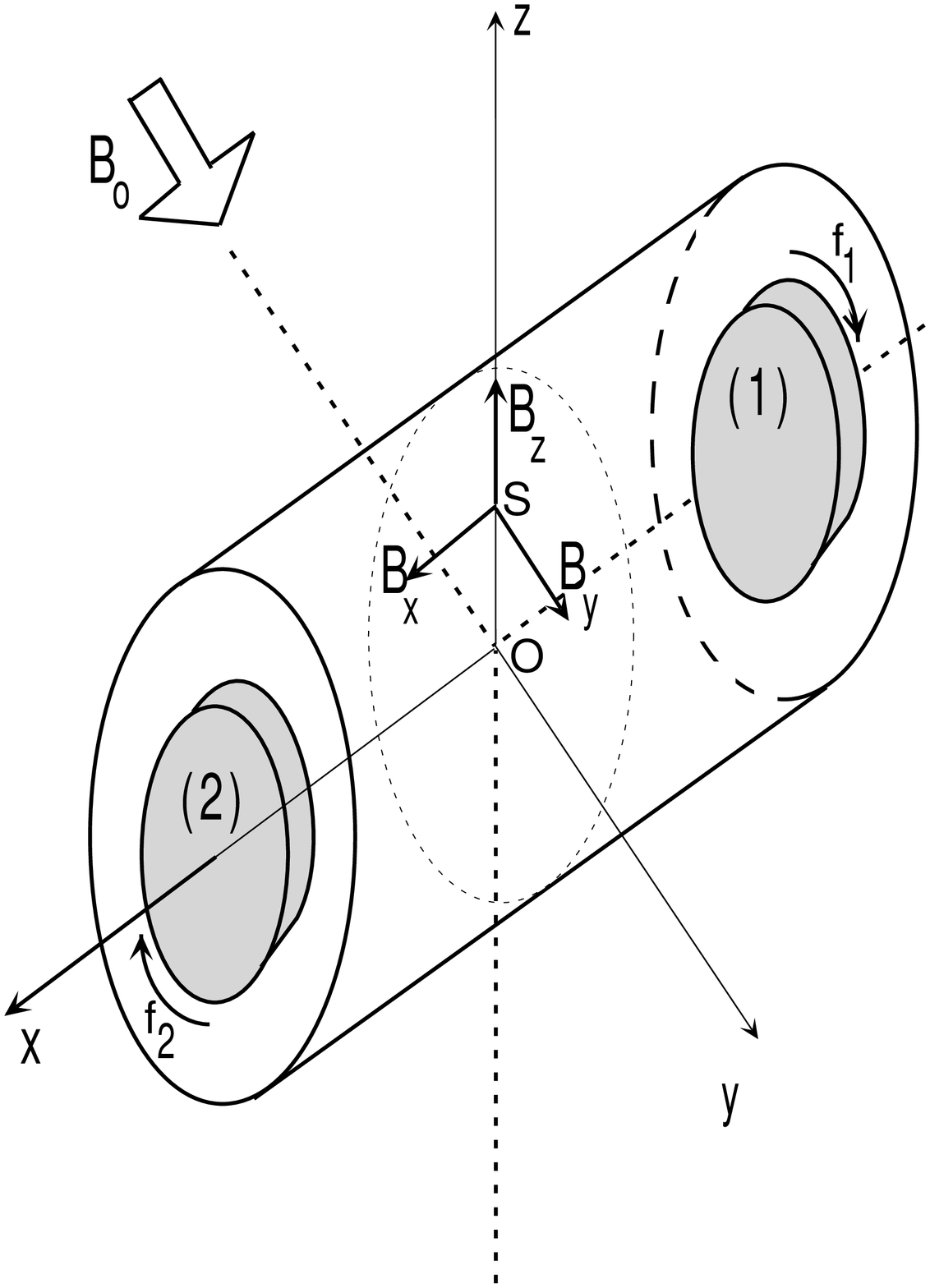}}
\caption{Geometry of the experimental set-up. The flow is generated by
rotating only one disk either at position (1) or (2). The magnetic
field is measured at position $S$. 
\label{experiment}}
\end{figure}

We have measured the induced magnetic field $\vec{B}$ generated by
a turbulent von K\'arm\'an swirling flow of liquid sodium submitted 
to a transverse external magnetic field $\vec{B}_0$ (see Fig.
\ref{experiment}).  
The sodium flow is operated in a loop that has 
been described elsewhere together with the  details of the experimental 
set-up \cite{vks01}. 
The flow is driven by rotating one of the two disks of radius $R$
located at position (1) or (2) in a cylindrical vessel, $40$ cm in inner
diameter and $40$ cm in length. In most experiments presented here, 
we use a disk of radius $R=150$ mm, fitted with 8 straight blades of
height $h=10$ mm driven at
a rotation frequency up to $f=30$ Hz. Four baffles, $20$ mm in height, have
been mounted on the cylindrical vessel inner wall, parallel to its axis.
A turbulent swirling flow with an integral Reynolds number, $Re=2\pi
R^2 f/\nu$, up
to  $3 \times 10^6$ is driven by the rotating disk. The mean flow has
the following characteristics: the fluid is
ejected radially outward by the disk; this drives an axial flow toward
the disk along its axis and a recirculation in the opposite direction
along the cylinder lateral boundary. The baffles inhibit the azimuthal
velocity of the recirculating flow and thus prevent a global rotation
of the fluid.  In some experiments,  we have used a disk of radius
$R=190$ mm, fitted with 16 curved blades of height $h=40$ mm, with or
without the lateral baffles in order to  observe the effect of a
stronger azimuthal flow. 

Two Helmholtz coils generate a magnetic field $\vec{B}_0$, perpendicular to the
cylinder axis (see Fig. \ref{experiment}).  The three components
of the field induced by the flow are measured with a 3D Hall probe,
located $180$ mm away from the disk in
the plane perpendicular to  $\vec{B}_0$ and containing the rotation
axis. The probe distance from the rotation axis is adjustable 
($z= 42, 100, 150$ mm).

The equations governing the magnetic field 
$\vec{B}_0 + \vec{B}(\vec {r}, t) $, where
$\vec{B}(\vec {r}, t)$ is the magnetic field generated by the flow
in the presence of the applied field $\vec{B}_0$, are in the MHD
approximation,

\begin{equation}
\vec{\nabla} \cdot \vec{B} = 0,
\label{divb}
\end{equation}
\begin{equation}
\frac{\partial{\vec {B}}}{\partial{t}} = \vec{\nabla} \times \left(\vec{V}
\times (\vec {B} + \vec {B}_0)\right) + {1 \over \mu_0 \sigma} \Delta \vec {B},
\label{ind}
\end{equation}
where $\vec{V}(\vec {r}, t)$ is the velocity field, $\mu_0$ is the
magnetic permeability of vacuum, and
$\sigma$ is the fluid electric conductivity.
 
The reaction of the magnetic field on the flow is characterized by
the ratio of the Lorentz force to the characteristic pressure forces
driving the flow. This is measured by the interaction parameter, 
$N = B_0^2 / \rho \mu_0 U^2$, where $\rho$ is the fluid density and
$U$ is the characteristic velocity of the solid
boundaries driving the fluid motion. The maximum field amplitude being $B_0 = 12$ G, $N$ is in the range
$10^{-5}-10^{-3}$, thus the effect of the magnetic field on the flow is
negligible in our experiments. This has been checked directly by measuring the
induced magnetic field as a function of the applied one at a constant driving
of the flow. We calculate the mean induced field
$\langle \vec{B} \rangle$ where $\langle \cdot \rangle$ stands for average
in time, as well as its $rms$ fluctuations in time,
$\vec{B}_{rms}$. Both vary linearly with $B_0$, thus showing that the
modification of the velocity field $\vec{V}$ in Eq. (\ref{ind}) can be neglected
 when $B_0$ is increased \cite{vks01}.
Thus, the only relevant dimensionless parameter of our experiments is the
magnetic Reynolds number, $R_m = \mu_0 \sigma R U=2 \pi \mu_0 \sigma
R^2 f$, which is proportional to the rotation frequency $f$ and has
been varied up to 40 for radius of the disks $R=150$ mm (respectively
$55$ for $R=190$ mm).

\begin{figure}
\centerline{\includegraphics[width=15cm]{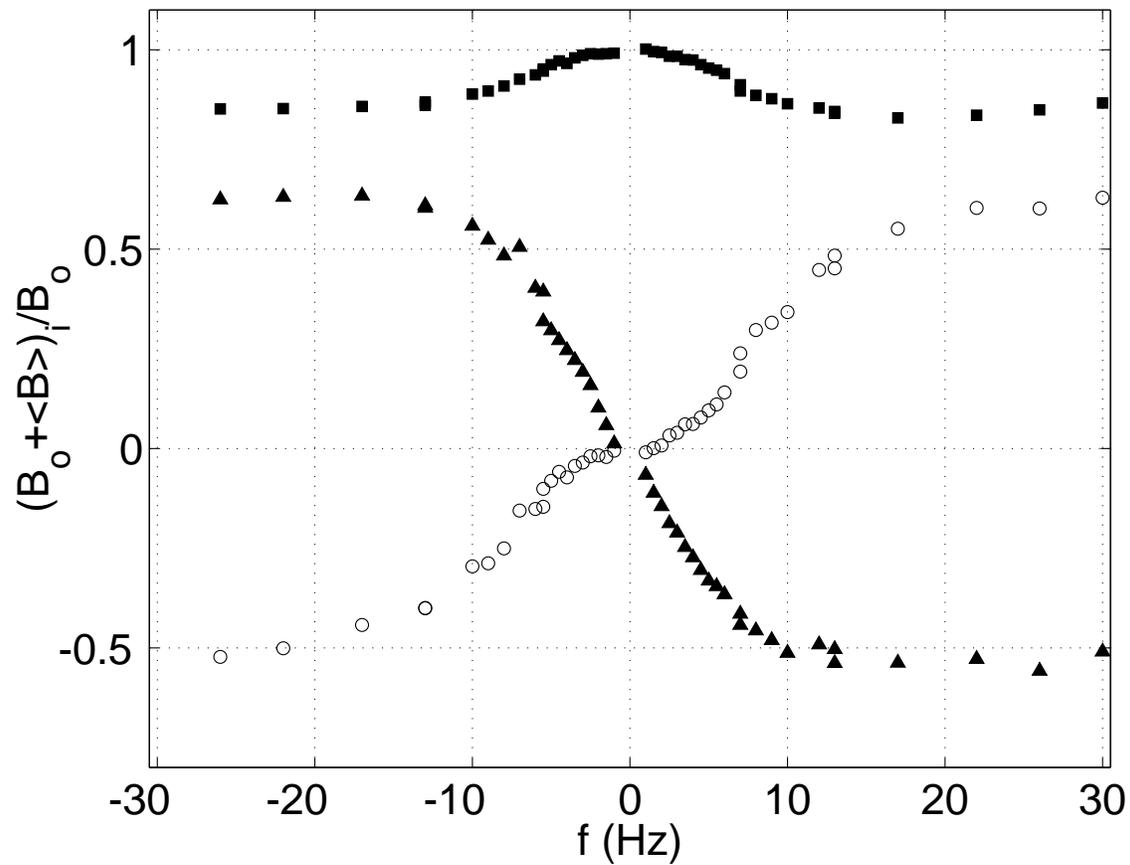}}
\vskip 0.4 truecm
\caption{Components of the total mean magnetic field as a function of the rotation
frequency of disk (2). The disk radius is $R=150$ mm with straight
blades. Four baffles are mounted on the
inner wall of the cylindrical vessel. The magnetic field is measured at
$z=100$ mm. (($o$) $=\frac{\langle B_x
\rangle}{B_o}$,   ($\blacksquare$)$= \frac{B_o + \langle B_y
\rangle}{B_o}$,($\blacktriangle$)$=\frac{\langle B_z
\rangle}{B_o}$)
\label{newinduced}}
\end{figure}

The three components of the mean magnetic field $\vec{B}_0 + \langle
\vec{B}(\vec {r}) \rangle$, at $z = 100$ mm above the rotation axis,
are displayed in Fig. \ref{newinduced} as a function of the rotation
frequency. We observe that when the rotation of the disk is reversed, $f 
\rightarrow -f$, we
approximately get
$\left(\langle B_x \rangle, \langle B_y \rangle, \langle B_z \rangle
\right) \rightarrow \left(- \langle B_x \rangle, \langle B_y \rangle,
- \langle B_z \rangle \right)$. When disk (2) is rotated instead of (1)
but keeping $f$ unchanged, we get
$\left(\langle B_x \rangle, \langle B_y \rangle, \langle B_z \rangle
\right) \rightarrow \left(- \langle B_x \rangle, \langle B_y \rangle,
\langle B_z \rangle \right)$ (note that the measurements of $\vec{B}$
are performed in the mid-plane between the two disks).
Assuming that the swirling flow has not broken the symmetries of the driving 
configuration, the above transformations of the field components can be
understood using the following symmetry transformations:

- (i) the symmetry with respect to the vertical plane perpendicular to
$\vec{B}_0$, $x0z$, shows that
if the disk is rotated in the opposite way, $f \rightarrow - f$,
we get $\left(\langle B_x \rangle, \langle B_y \rangle, \langle B_z \rangle
\right) \rightarrow \left(- \langle B_x \rangle, \langle B_y \rangle, -
\langle B_z \rangle \right)$ ($\vec{B}$ is a pseudovector).

- (ii) The symmetry with respect to the vertical plane parallel to
$\vec{B}_0$, $y0z$, followed by the transformation $\vec{B}_0 \rightarrow
- \vec{B}_0$, shows that when we rotate disk (2) instead of disk (1)
without changing the sign of $f$, we get $\left(\langle B_x \rangle,
\langle B_y \rangle, \langle B_z \rangle \right) \rightarrow 
\left(- \langle B_x \rangle, \langle B_y \rangle, \langle B_z \rangle \right)$. 

The induced field component $\langle B_y \rangle$ is opposed to
$\vec{B}_0$ and increases in
amplitude, thus the total field along $\vec{B}_0$ decreases as $R_m$ is
increased.  This expulsion of a transverse magnetic field from eddies is well
documented, both theoretically \cite{par66,wei66} and experimentally
\cite{odi00}. The expulsion is stronger close to the axis of the cylinder
($z=42$ mm). On the contrary, closer to the cylinder lateral boundary
($z=150$ mm), the field increases with $R_m$. Thus, the field 
is expelled from the core of the swirling flow and concentrates at its 
periphery.
 
\begin{figure}
\centerline{\includegraphics[width=15cm]{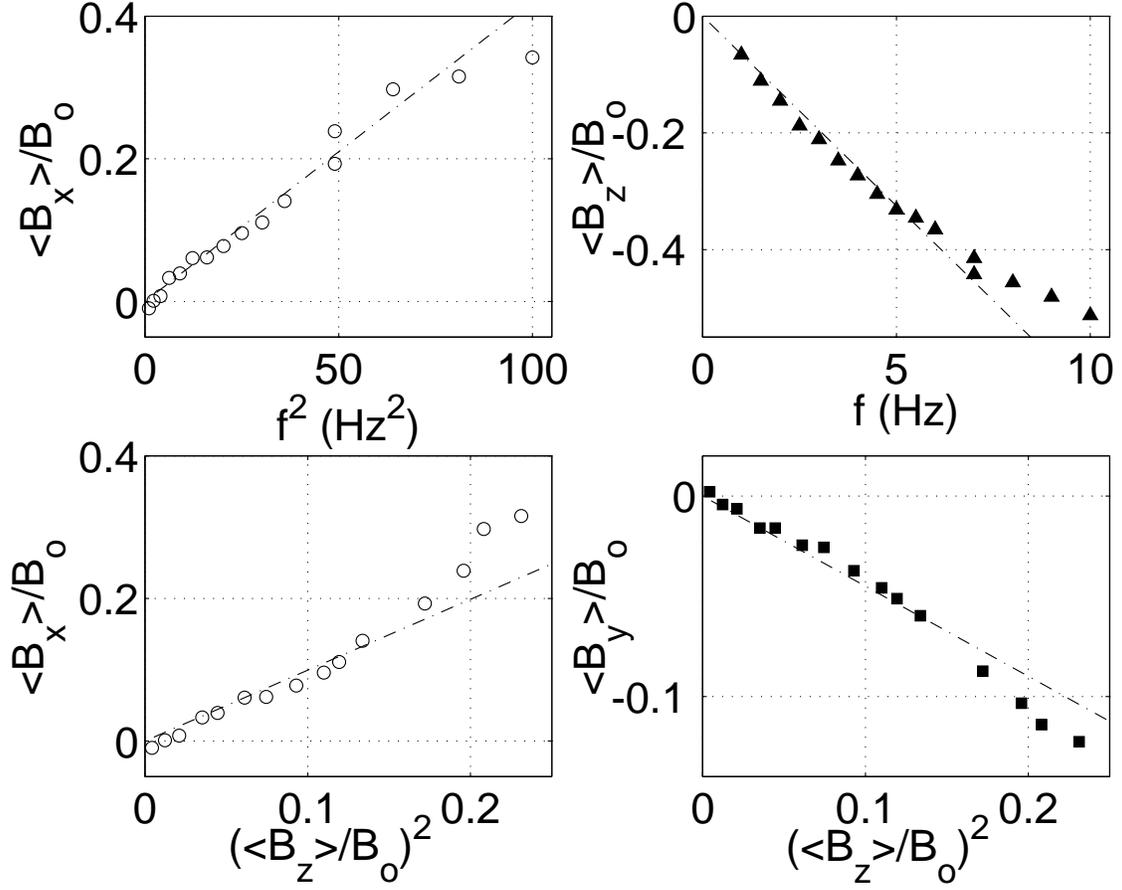}}
\vskip 0.4 truecm
\caption{(a-b): Axial and vertical mean components of the induced magnetic field as
a function of the rotation frequency. (c-d): Axial and transverse mean components of
the induced  magnetic field as a function of the square of the
vertical one, for $f < 10\, Hz$. Same
experimental configuration as in Fig. \ref{newinduced}.
\label{newscalings}}
\end{figure}

The components of the field induced perpendicular to $\vec{B}_0$ both
increase in amplitude from zero, reach a maximum and then saturate when $R_m$ is increased
further.  As shown in Fig.
\ref{newscalings}, these two components do not scale in the same way at small
rotation frequency, {\it i.e.} at small $R_m$. The amplitude of the vertical
component
$\langle B_z \rangle$ increases linearly  whereas the axial component
$\langle B_x \rangle$ increases quadratically with $R_m$. Indeed, at
the location of the measurements, we observe $\langle B_x \rangle \propto 
\langle B_z \rangle^2$ and $\langle B_y \rangle \propto 
\langle B_z \rangle^2$, roughly up to $f=10$ Hz (see Fig. \ref{newscalings}). 

Writing $\vec{B}(\vec {r}, t) = \langle \vec{B}(\vec {r}) \rangle + \vec{b}
(\vec {r}, t)$, 
and similarly for $\vec{V}$, we get from Eq. (\ref{ind}) for the mean induced field
\begin{equation}
- {1 \over \mu_0 \sigma} \Delta \vec {\langle B \rangle} =
\vec{\nabla} \times \left( \vec{\langle V \rangle} \times \vec {B}_0 +
\vec{\langle V \rangle} \times \vec{\langle B \rangle} +
\langle \vec{v} \times \vec{b} \rangle \right).
\label{indmean}
\end{equation}
When the magnetic Reynolds number is small, the first source term on the right
hand side of Eq. (\ref{indmean}) is the dominant term and we get for each component
of the mean induced field $\langle B_i \rangle \propto R_m B_0$. However,
both the expulsion of a transverse field from a rotating eddy and the
$\alpha$-effect, {\it i.e.} the generation of a current parallel to an
applied field by a cyclonic eddy, cannot be described at this level and
involve the nonlinear source terms  of Eq. (\ref{indmean}).

Indeed, keeping only the contribution of the first term on the right
hand side of equation (\ref{indmean}) gives for $\langle B_x \rangle$
\begin{equation}
- {1 \over \mu_0 \sigma} \Delta  {\langle B_x \rangle} =
{B}_0 \frac{\partial \langle V_x \rangle}{\partial y}\, .
\label{indlin}
\end{equation}
The source term being antisymmetric with respect to the vertical plane $x0z$,
we have for the linear response $\langle B_x(x, y=0, z) \rangle=0$.

The generation of $B_x$ is the consequence of the $\alpha$-effect,
{\it i.e.} of the
generation of a current parallel to $\vec{B}_0$. Indeed, at leading order in $R_m$,  
$\langle B_x \rangle$ increases like $R_m^2$ and its sign with respect
to the sign  of
$B_0$ is determined by the flow helicity, $h = \overline{\vec{V}\cdot
(\vec{\nabla} \times \vec{V})}$ where the overbar stands for the
spatial average. One can easily check that $\langle B_x \rangle$
changes sign under any symmetry with respect to a plane containing the
rotation axis, just as does the pseudoscalar $h$.

For fixed $R_m$, $\langle B_x \rangle$
increases with the distance to the cylinder axis in the range $42 < z < 150$ mm.
We can show that it should vanish for $z = 0$: indeed, the rotation of angle
$\pi$ around the $x$-axis followed by the transformation 
$\vec{B}_0  \rightarrow - \vec{B}_0$ which implies $\vec{B}
\rightarrow -  \vec{B}$, gives
$\langle B_x (x, 0, z) \rangle =  - \langle B_x (x, 0, -z) \rangle$.

\begin{figure}
\centerline{\includegraphics[width=15cm]{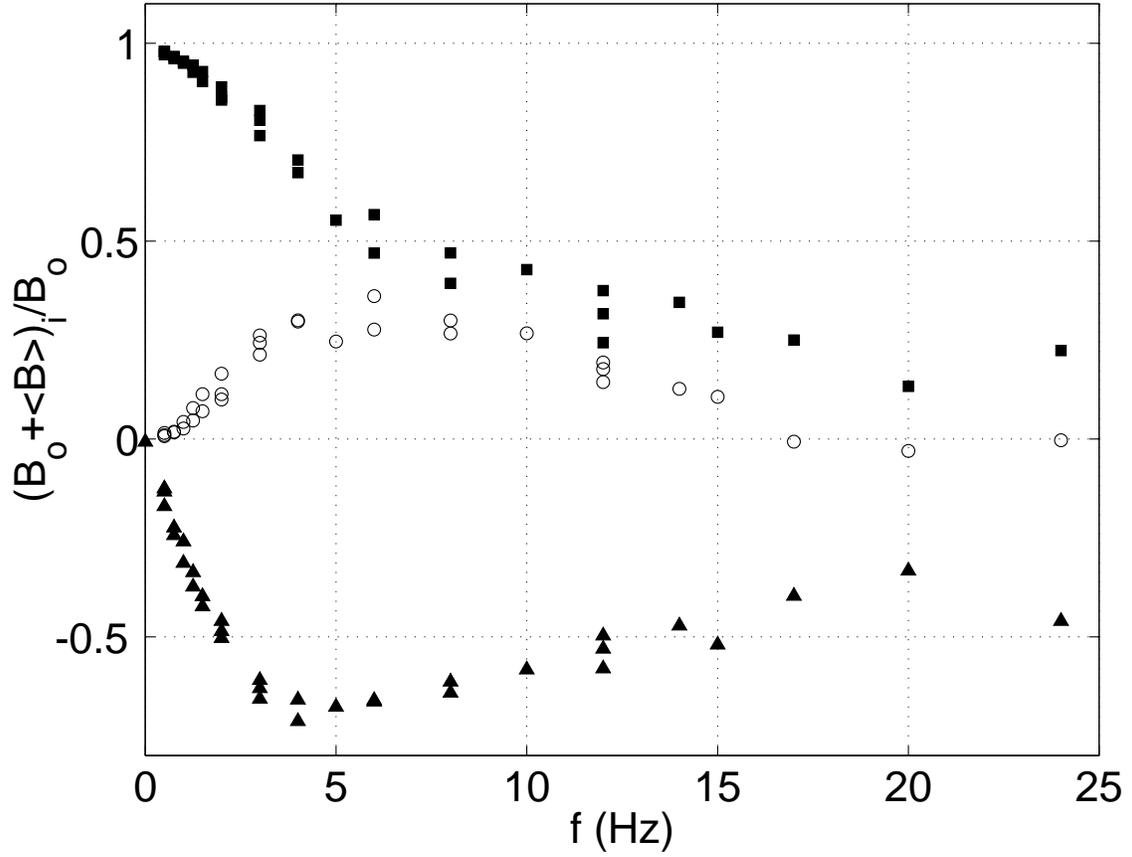}}
\vskip 0.4 truecm
\caption{Components of the total mean magnetic field as a function of
the rotation frequency of the disk (2). The disk radius is
$R=190$ mm with curved blades. There are no baffles on the inner wall of the cylindrical
vessel.  The magnetic field is measured at
$z=100$ mm. (($o$) $=\frac{\langle B_x
\rangle}{B_o}$,   ($\blacksquare$)$= \frac{B_o + \langle B_y
\rangle}{B_o}$,($\blacktriangle$)$=\frac{\langle B_z
\rangle}{B_o}$)
\label{induced}}
\end{figure}

When $R_m$ is increased $\langle B_x \rangle$ seems to saturate for $z=100$ mm
(see Fig.\ref{newinduced}). Closer to the rotation axis ($z = 42$ mm), it reaches
a maximum and then decreases roughly to zero when $f$ is increased up to $30$ Hz.
This is due to the expulsion of the applied magnetic field $\vec{B}_0$ from the core 
of the swirling flow. Indeed, when the baffles are removed from the cylindrical 
vessel inner wall, global rotation  of the flow is no longer inhibited, and   
$\langle B_x \rangle$ measured $100$ mm away from the rotation axis,  decreases 
to zero at large $R_m$ (see Fig. \ref{induced} and compare with
Fig. \ref{newinduced} where $\langle B_x \rangle$ stays finite at large $R_m$). Consequently, we observe that 
the $\alpha$-effect decays when $R_m$ is too large, or more precisely when
the  magnetic Reynolds number corresponding to the azimuthal flow is too
large, because of the transverse field expulsion from the cyclonic eddy. A
similar effect has been  recently computed by R\"adler et al. in the case of the
Roberts flow \cite{rad97}. They have shown by computing terms higher than the second
order in $R_m$, that  the $\alpha$-effect reaches a maximum and then tends to zero
when $R_m$ is increased further (compare their Fig. 3 with our
Fig. \ref{induced}).   Our measurements are the first experimental
demonstration that  for large $R_m$  the $\alpha$-effect can vanish  due to the geometry of the flow (i.e. when the
amplitude of the azimuthal component of the swirling flow is too large
compared to the axial
component).

Using dimensional analysis, we can write
\begin{equation}
\langle B_x (\vec{r}) \rangle = B_0 \, {\cal F}(\vec{r}, R_m, P_m, N),
\label{bind}
\end{equation}
where $P_m = \mu_0 \sigma \nu$ is the magnetic Prandtl number ($\nu$ is the
kinematic viscosity). As said above, for $B_0$ small enough, ${\cal F}$ does not depend on
the interaction parameter $N$. 
The only previous experimental study of the $\alpha$-effect \cite{ste68} considers
a constrained flow configuration in a range of $B_0$ for which a dependence on $N$ but no dependence on $R_m$ have been found. The dependence on $N$
can give insights on the nonlinear saturation mechanism of a dynamo generated via
the $\alpha$-effect. On the contrary, the dependence on $R_m$ determines the dependence of
the linear growth rate of a dynamo generated via the $\alpha$-effect. The decay of the $\alpha$-effect for large
$R_m$ reported here,  gives a possible mechanism for a ``slow dynamo", i.e. a dynamo with 
a growth rate that decreases at large $R_m$. 
Finally, the dependence on $P_m$, or equivalently on the
Reynolds number of the flow, cannot be determined from our measurements. 
In other words, we do not know the contribution of the turbulent fluctuations
to the measured $\alpha$-effect, i.e. the relative contribution of the two
nonlinear source terms $\vec{\nabla} \times \left(\vec{\langle V \rangle} \times
\vec{\langle B \rangle}\right)$ and  $\vec{\nabla} \times \left(\langle \vec{v}
\times \vec{b} \rangle \right)$ in equation (\ref{indmean}). Both the
velocity field, measured in water \cite{vks01}, and
the magnetic field display large fluctuations (roughly $20\%$), but we cannot evaluate
$\langle \vec{v} \times
\vec{b} \rangle$. It would be of  great interest to develop a device for
simultaneous measurements of magnetic and velocity fields, or to perform
experiments with different liquid metals in order to quantify the effect of $P_m$.
Indeed, the effect of $P_m$ on the induced fields can give insights in the problem
of the dependence of an $\alpha$-dynamo threshold,
$R_m^{c}$, on $P_m$ or equivalently on the Reynolds number of the flow. 
The behavior of $R_m^{c}$ in the limit of large Reynolds number is an open
problem of kinematic dynamo theory and is of prime experimental and theoretical interest.   

We gratefully acknowledge the assistance of J.-B. Luciani and
M. Moulin and the financial support of the french institutions:
Direction des Sciences de la Mati\`ere and Direction de l'Energie Nucl\'eaire
of CEA, Minist\`ere de la Recherche and
Centre National de Recherche Scientifique. J. Burguete was supported
by post-doctoral grant No. PB98-0208 from ministerio de Educacion y
Ciencias (Spain) while at CEA-Saclay.


\vfill
\eject
\vfill
\eject

\vfill
\eject

\vfill
\eject



\end{document}